\documentclass[9pt]{article}
\usepackage{spconf, amsmath,graphicx}

\usepackage{mathrsfs, amsfonts, amsmath, dsfont}
\usepackage{graphicx}
\usepackage{color}
\usepackage{booktabs}
\usepackage{stfloats}
\usepackage{subfigure}
\usepackage{multirow}
\usepackage{tikz, pgfplots}
\usetikzlibrary{shapes,arrows}
\tikzset{%
	block/.style    = {draw, thick, rectangle, minimum height = 3em,
		minimum width = 3em},
	sum/.style      = {draw, circle, node distance = 2cm}, 
	multi/.style      = {draw, circle, node distance = 2cm}, 
	input/.style    = {coordinate}, 
	output/.style   = {coordinate} 
}
\usepackage{subfigure}
\usepackage{longtable}

\title{An Efficient Modal-based Approach Towards Guzheng Sound Synthesis}
\name{Enda Zhang$^{\star}$ \qquad Gopal Gupta$^{\star}$ \qquad Charles Greif \qquad Andrew Paplinski$^{\star}$}

			\address{$^{\star}$ Monash University }

\begin{document}

\ninept
\maketitle
\begin{abstract}

	Encouraged by recent interest in traditional Chinese instruments this work proposes a computational sound synthesis model for a traditional Chinese instrument, the guzheng. Digital waveguide model and modal synthesis are the most popular physical modelling methods. However, the excitation signal is usually over-simplified and the design of some digital filters is inadequate for real-time synthesis. We propose an accurate and efficient approach to synthesise the string signal of the guzheng. Our proposed model enables accurate parametric control of the pluck because the parameters are based on physical model of the instrument. The sound of the guzheng is then computed as the convolution of the synthesised string signal and the impulse response of the resonant body. As proof of concept, the 21st string of the guzheng is synthesised using the proposed model. The sound computed using our model shows improved accuracy with an acceptable computational complexity.
\end{abstract}


\section{Introduction}
\label{sec:intro}

The \emph{guzheng} is a traditional Chinese plucked string musical instrument. It is widely agreed that the \emph{guzheng} firstly appeared during the Qin Dynasty of Imperial China (221-206 BCE), and it evolves from another traditional Chinese musical instrument \emph{se}. More strings have been added  and its structure has been evolving gradually to the modern \emph{guzheng}.

\begin{figure}[tb!]
	\centering
	\includegraphics[width=0.8\linewidth]{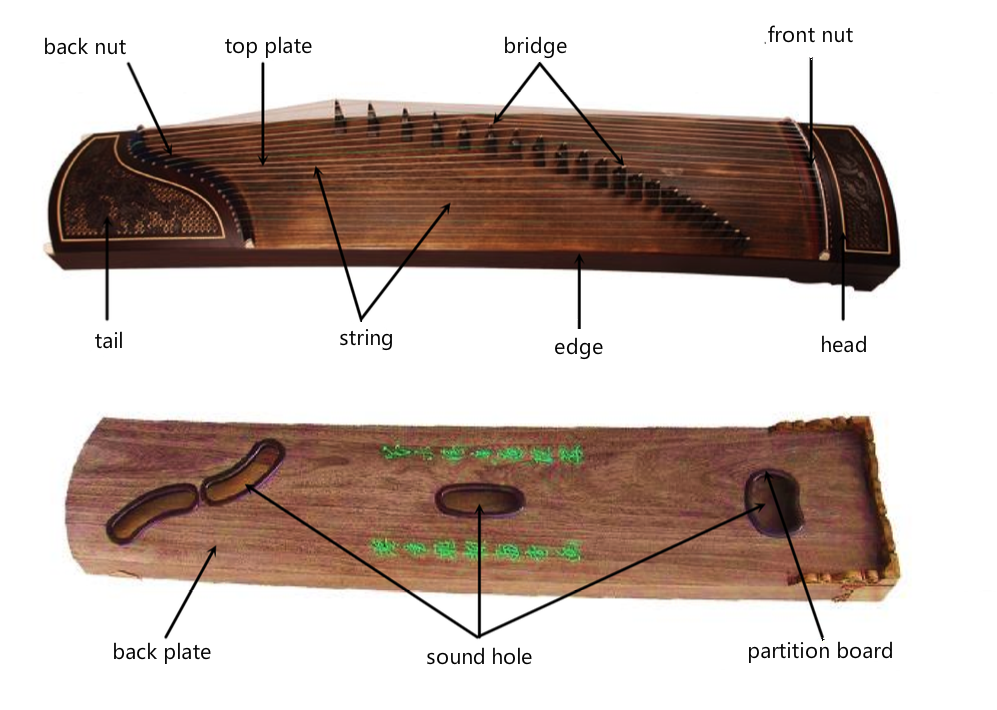}
	
	\caption{The structure of the \emph{guzheng}. Source: \cite[Figure 2-1]{Deng2015thesis}.}
	\label{fig:Guzheng-structure}
\end{figure}

The \emph{guzheng} has 21 strings which are made of either nylon or steel. The strings are mounted on the top plate of the body through the front nut and back nut. The vibration of the string is transmitted to the body through a bridge, which is located in a line on the top plate (as shown in Fig. \ref{fig:Guzheng-structure}). The length of the string and therefore the fundamental frequency depends on both the tension of the string and the position of the bridge. The body of the \emph{guzheng} is a sound box designed to be curved to balance the tension of strings. There are three or four soundholes located on the back plate to couple the sound of the \emph{guzheng} to the air \cite{Deng2015dynamic} \cite{Deng2015thesis}.

When playing the \emph{guzheng}, the string segment between the bridge and the front nut (right-half of the string) is to be plucked, while the other segment (left-half of the string) is used for pitch variation by pressing the string. Players wear plastic nails on their right hand when playing, in order to create sharper and more dynamic sound.


To date, very little research of the sound and acoustics of the \emph{guzheng} has been published. Xiaowei Deng, a researcher from Shanghai Jiao Tong University, completed his master thesis based on a project of the acoustics of the \emph{guzheng}, which was set up with Shanghai National Musical Instrument Factory. Deng \cite{Deng2016simulation} both studied the mechanical structure  and performed lots of measurements on the physical properties of the \emph{guzheng}. Also, a set of partial differential equations was proposed to synthesise the vibration signal and the sound of the 21st string \cite{Deng2015dynamic} \cite{Deng2015thesis}.

\emph{Digital Waveguide} (DWG) model and \emph{Modal-based Synthesis} are two most popular methods to efficiently synthesise the sound of a string instrument. Digital Waveguide Model (DWG), proposed by Julius O. Smith, can be interpreted as an efficient implementation of d'Alembert's formula using delay lines in the one-dimensional case \cite{smith1993efficient}. Low-pass filters and all-pass filters could be added to the delay line to simulate the damping effect and the stiffness of the string \cite{smith1993efficient} \cite{smith2010physical}. The input of DWG model can be a wavetable or an excitation signal generated by a parametric synthesis model. Parametric synthesis frameworks have been applied to the synthesis of the acoustic guitar \cite{valimaki1998development}, electric guitar \cite{lindroos2011parametric} and the clavinet \cite{gabrielli2013digital}. However, it is challenging to accurately synthesise an excitation signal with control parameters since it mimics the effect of a pluck. It would be more intuitive to consider a ``physical'' way to produce the excitation signal which comes directly from a physical model. 


Another approach, modal synthesis, produces the sound as the summation of exponentially decaying sinusoids, which represent the eigenmodes of a linear system.
Trautmann and Rabenstein proposed and applied functional transformation methods (FTM) to convert partial differential equations into modal representations. As a proof of the concept, they applied FTM to the extended 1-dimensional wave equation for the synthesis of the classic guitar \cite{rabenstein2003digital}. FTM is usually computationally expensive when the number of modes is large. In 2010, Bank et al. \cite{bank2010modal} proposed a modal-based piano synthesiser where all the hammer, string and soundboard are modelled by the modal synthesis. After optimisation, the model is guaranteed to run in real-time.

To reduce the computation complexity of FTM, a combination method was proposed by Trautmann et al. which calculates the parameters in DWG using the result of FTM instead of estimating from measurement \cite{trautmann2002combining}. Rectangular functions were used as the excitation signal of the DWG. The model is more efficient than using FTM directly while keeps the precision at a desirable level. However, using a rectangular function as the excitation signal is inherently inaccurate since the vibration after propagating on the string has a more structured waveform. The rectangular excitation signal does not allow effective parametric control of a physical model. Furthermore, the original design of the dispersion filter is not a closed form, therefore is hard to implement in real-time.

To the best of our knowledge, Deng's approach is the first computational model for the \emph{guzheng} synthesis. As Deng modelled the string using a 1-dimensional ideal wave equation, the vibration signal was produced based on the physical parameters only. Therefore the model can be adjusted easily by the physical parameters and the synthesised sound is relatively accurate. However, as sound synthesis was not Deng's main focus, his approach has some drawbacks and therefore needs to be improved. First, his model synthesises the sound based on numerical analysis of partial differential equations. It is hard to develop a sound synthesiser using his approach as his synthesis model cannot run in real-time. Second, the damping effect and the stiffness of the string were not included in his model. Third, the plucking was modelled by the initial displacement and initial velocity, which are caused by the plucking force and needs further measurement.

Some similar string instruments have been modelled in previous studies.
Kantele is a traditional Finish musical instrument, which has a similar acoustic structure as the zither and the \emph{guzheng} \cite{erkut2002acoustical}. Guqin, as a similar instrument compared to the \emph{guzheng}, has been thoroughly studied in its acoustics \cite{waltham2016acoustical}. Both Guqin \cite{penttinen2006model} and Kantele \cite{karjalainen1993analysis} have been synthesised using DWG model and pre-recorded excitation signals. 

Consequently, in this paper, we propose a computational synthesis model for the \emph{guzheng}. The model is as efficient as typical DWG-based models while keeps the accuracy at a desirable level by adopting modal synthesis. The stiffness of the string and pitch varying are also considered in our proposed model.
The novelty of this work includes:
\begin{itemize}
	\item Using modal synthesis to produce the excitation signal to achieve high accuracy
	\item Allowing parametric control of the pluck using parameters and physical characteristics of the string
	\item Enabling real-time synthesis by changing the design of the dispersion filter
	\item Applying our proposed method to the \emph{guzheng}, of which a real-time synthesiser is yet to be created.
\end{itemize}

The method for the estimation of parameters and the design of each part of the model are shown in Section \ref{sec:method}. The results and discussion are in Section \ref{sec:results} and Section \ref{sec:discussion}, respectively. Finally, the conclusions and discussion are in Section \ref{sec:conclusion}.

\section{Proposed Method}
\label{sec:method}

The model consists of three main parts: modal synthesis, Digital Waveguide Model and the convolution. Using functional transformation method, modal synthesis part produces an excitation signal\footnote{Note that different from the terminology researchers use in DWG, we name the output of modal synthesis as \emph{excitation signal} and the input of modal synthesis part as \emph{excitation force signal}.} for DWG part, where the physical and damping parameters are the control parameters. The damping parameters, describing the damping effect of the particular string, can be estimated empirically or retrieved from a recorded vibration signal of the string. Next, the DWG model processes the excitation signal by a delay line and multiple digital filters. Such filters handles both the tone and the pitch of the synthesised sound. The final sound of the \emph{guzheng} is produced by the convolution of the string signal and the impulse response of the body. 

The detail of each part will be discussed in the following sections. The process and calculation of modal synthesis part are shown in section \ref{sec:method-ftm}. The method used for estimating the damping parameters in the PDE will be provided in section \ref{sec:method-damping-param}. Details are given in the following subsections. The design of the low-pass filter and all-pass dispersion filter used in DWG part is listed in section \ref{sec:method-lp} and \ref{sec:method-dispersion}, respectively. The consideration for fine-tuning is discussed in section \ref{sec:method-tuning}.

\subsection{Construction of Excitation Signal}
\label{sec:method-ftm}
As introduced previously, we will feed a short clip of wave signal into one-dimensional digital waveguide model. To achieve high accuracy, we synthesise the wave signal by FTM as a modal approach. Modal synthesis has been widely used in the synthesis of string instruments, and the detailed description can be found in many research works and textbooks, especially in \cite{rabenstein2003digital} \cite{trautmann2003digital} \cite{bank2010modal}. 

Guzheng has thicker strings compared to popular string instrument such as guitar. Therefore, in our case, the stiffness and damping effect on the string of the \emph{guzheng} should not be neglected \cite{Deng2015dynamic}. We take the dispersion and damping factors into account and consider the following extended one-dimensional wave equation
\begin{equation}
	\label{equ:modified-FTM-wave-equation}
	\left\{
	\begin{aligned}
		 & \epsilon \frac{\partial^2 y}{\partial t^2} + EI \frac{\partial^4 y}{\partial x^4} - T_s \frac{\partial^2 y}{\partial x^2} + d_1 \frac{\partial y}{\partial t} + d_3 \frac{\partial^3 y}{\partial t \partial x^2} = \mathcal{F}_e(x,t), \\
		 & y(x\;, 0) = 0,~\dot{y}\;\,(x\;, 0) = 0, ~~~~~~~~~~~~~~~~~~~~0 \leq x \leq l,                                                                                                                                                 \\
		 & y(x_b, t) = 0,~y''(x_b, t) = 0, ~~~~~~~~~~~~~~~~~~~~x_b \in \{0,~l\}.
	\end{aligned}\right.
\end{equation}
where $x \in [0,~l]$ is the spatial variable, $t \in [0,+\infty)$ is the temporal variable, $y(x, t)$ is the displacement of the string, $\epsilon$ is the linear density, $T_s$ is the tension of the string, $E$ is the Young's modulus of the string and $I$ is the moment of inertia. $d_1$ and $d_3$ are the frequency dependent and independent decay parameter, respectively. The method for estimating $d_1$ and $d_3$ will be discussed in section \ref{sec:method-damping-param}. 

On the \emph{guzheng}, the pluck point is fixed when plucking, hence we can separate the temporal and spatial variables in general cases. In practive, the excitation force can be simplified as $\mathcal{F}_e(x, t) = \mathcal{F}_{e1}(x)\mathcal{F}_{e2}(t)$. These two parts can be further expressed as
$\mathcal{F}_{e1}(x)  = f \delta(x-x_e)$ and 
$\mathcal{F}_{e2}(t)  = \mathds{1}_{[0, t_p]}$ 
where $f$ is the magnitude of the excitation function, $x_e$ is the pluck point, $t_p$ is the time interval of the pluck and $\delta(x-x_e)$ is a pulse at $x = x_e$. The notation $\mathds{1}_{[0, t_p]}$ represents an rectangular function, which takes $1$ when $x \in [0, t_p]$ and $0$ otherwise. It is a reasonable approximation because the pluck point is always fixed when the string is plucked, and only the excitation force varies with time. Also, we assume the displacement $y(x, t)$ is sufficiently smooth.

By applying a sequence of functional transforms to Eq.\ref{equ:modified-FTM-wave-equation}, the PDE is transformed into a multi-dimensional discrete transfer function \cite{rabenstein2003digital}. The discrete transfer function consists multiple second-order IIR filters which represent the modes in the excitation signal. Let $\mu$ denote the mode number and $x_a$ denote the point on string we calculate, then the discrete transfer function can be represented as
\begin{equation}
	H(x,z) = \sum_{\mu} a(\mu, x_a) \frac{b_1(\mu)z^{-1}}{1 + c_1(\mu)z^{-1} + c_0(\mu)z^{-2}}.
	\label{equ:FTM-Hd}
\end{equation}
The coefficients of the IIR filters are calculated as
$$
	\begin{aligned}
		a(\mu, x_a) & = \frac{1}{2}fl\sin(\frac{\mu \pi x_a}{l})\sin(\frac{\mu \pi x_e}{l}), \\
		b_1(\mu)  & = \frac{1}{\epsilon \omega_\mu} \sin(\omega_\mu T),                    \\
		c_1(\mu)  & = -2\exp(\sigma_\mu T)\cos(\omega_\mu T),                              \\
		c_0(\mu)  & = \exp(2\sigma_\mu T),
	\end{aligned}
$$
where
$$
	\begin{aligned}
		\gamma_\mu   & = \frac{\mu\pi}{l},                                                                                                                                     \\
		\sigma_\mu   & = \frac{1}{2\epsilon}(d_3 \gamma_\mu^2 - d_1),                                                                                                                        \\
		\omega_\mu^2 & = \left(\frac{4\epsilon EI - d_3^2}{4\epsilon^2}\right)\gamma_\mu^4 + \left(\frac{2\epsilon T_s + d_1 d_3}{2\epsilon^2}\right)\gamma_\mu^2 - \frac{d_1^2}{2\epsilon}.
	\end{aligned}
$$
In the formulae above, $T = 1/f_s$ denotes the sampling interval. Computationally, (\ref{equ:FTM-Hd}) can be implemented as the weighted average of the output of a large numbers of second-order IIR filters in parallel \cite{rabenstein2003digital}.

\subsection{Estimating Damping Parameters $d_1$ and $d_3$}
\label{sec:method-damping-param}

The damping parameters, controlling the decay rate of the string vibration, can be either estimated from a recorded string vibration signal or guestimated based on experience.  
If a pre-recorded string signal is available, the damping parameter $d_1$ and $d_3$ can be derived by linear regression.
\subsection{Design of the Low-pass Filter}
\label{sec:method-lp}

In our proposed model, the propagation of the wave on the string is modelled by a digital waveguide model. The damping effect is caused by the energy loss of the string in the air, and the transmission from string to bridge. 


As suggested by Rabenstein and Trautmann \cite{trautmann2003digital}, a one-pole filter can be designed based on the result from mathematical transforms to the original physical model. The filter coefficients are calculated by the physical parameters mentioned in (\ref{equ:FTM-Hd}) only. To be specific, 
$g = 1 - {\pi d_1}/{\epsilon \omega_1 }$, $a = -1 + c_0 - \sqrt{c_0^2 - 2c_0}$,
where $c_0 = -\frac{\epsilon l^2 \omega_1^3 T^2}{4pi^2 d_3}$.

\subsection{Design of the Dispersion Filter}
\label{sec:method-dispersion}

Similar to the damping effect, the dispersion effect of the string can be modelled by adding a dispersion filter into the loop of DWG model.
Usually the design of the dispersion filter requires a lot a computation and can hardly run in real-time. To enable on-line processing, Thiran filter can be used to model the dispersion effect of piano and other string instruments \cite{rauhala2006tunable}. The Thiran filter \cite{thiran1971recursive} is a second-order all-pass filter
$$
	A_{disp}(z) = \frac{a_2 + a_1 z^{-1} + z^{-2}}{1 + a_1 z^{-1} + a_2 z^{-2}}
$$
where
\begin{equation}
	a_k = (-1)^{k} \left(\begin{matrix}N\\k\end{matrix}\right) \prod_{n = 0}^{2} \frac{D-2+n}{D-2+k+n},~~~~k = 1,~2.
	\label{equ:Thiran-coef}
\end{equation}

We use four cascade Thiran filters as an all-pass dispersion filter to model the stiffness of the string of the \emph{guzheng}. Let $D_{disp}$ denote the parameter used for the dispersion filter design, the coefficient $D = D_{disp}$ is estimated based on the pitch of the sound and the physical parameters, including Young's modulus, diameter, tension and length of the string \cite{rauhala2006tunable}.

\subsection{Consideration for Tuning}
\label{sec:method-tuning}

To produce the sound with fundamental frequency $f_0$, a total delay of $f_s/f_0$ is needed in the DWG part to create the periodical signal. Therefore, except for an integer-size delay unit for coarse tuning, it is necessary to have a fractional delay for fine-tuning. The Thiran filters can create group phase delay of $D$ \cite{laakso1996splitting}, and the dispersion filter $A_{disp}(z)$ can create $4D_{disp}$ of delay.
Thus, the integer-size delay should take
$
	L = \lfloor {f_s}/{f_0} - 4D_{disp} \rfloor - 1
$
units, and the delay we need for fine tuning is calculated as    $D_{frac} = {f_s}/{f_0} - L - 4D_{disp}.$ The coefficients of $A_{frac}(z)$ are calculated by (\ref{equ:Thiran-coef}) with $D = D_{frac}$ \cite{laakso1996splitting} where $D_{frac}$ denote the parameter used for the design of the fraction delay filter.

\section{Results}
\label{sec:results}

\begin{table}[!tb]
\scriptsize
	\centering
	\caption{The coefficient for the filters used in the DWG part of the synthesis model.}
	\begin{tabular}{c|ccc}
		\hline \hline
		\multirow{2}{*}{$A_{d}(z)$}  & $D_d$     & $a_1$                & $a_2$     \\\cline{2-4}
		& $15.5236$ & $-1.6369$            & $0.6783$  \\
		\hline \hline
		\multirow{2}{*}{$A_{f}(z)$}  & $D_f$     & $a_1$                & $a_2$     \\\cline{2-4}
		& $1.8566$  & $0.1004$             & $-0.0112$ \\
		\hline \hline
		\multirow{2}{*}{$H_{lp}(z)$} & $g$       & $a$                  &           \\\cline{2-4}
		& $0.9985$  & $2.31\times 10^{-7}$ &           \\
		\hline \hline
	\end{tabular}
	
	\label{tab:filter-coef}
\end{table}

\begin{table}[!tb]
\scriptsize
	\centering
	\caption{The physical parameters of 21st string of \emph{guzheng} \cite{Deng2015thesis} \cite{Deng2016simulation}}.
	\begin{tabular}{ccc}
		\hline
		\hline
		Parameter  & Description        & Value                                \\
		\hline
		$\epsilon$ & Linear density     & $2.057 \cdot 10^{-2}~\text{kg/m}$    \\
		$E$        & Young's Modulus    & $220~\text{Gpa}$                     \\
		$d$        & Diameter of string & $0.6~\text{mm}$                      \\
		$I$        & Moment of inertia  & $2.545 \cdot 10^{-2}~ \text{mm}^{4}$ \\
		$l$        & Length of string   & $950~\text{mm}$                      \\
		$T_s$      & Tension of string  & $400~\text{N}$                       \\
		\hline
		\hline
	\end{tabular}

	\label{tab:physical-properties-Guzheng}
\end{table}

\begin{figure}[!tb]
	\centering
	\includegraphics[width=0.35\textwidth]{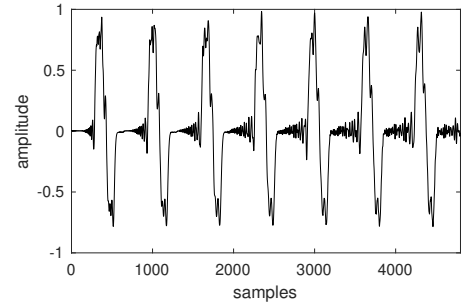}
	\caption{The string signal synthesised by proposed combination contains 4800 samples with sampling rate $f_s = 48k$Hz.}
	\label{fig:synthesised-string}
\end{figure}

We firstly estimate the damping parameters. The recorded string signal was obtained from Deng \cite{Deng2015thesis}, who measured physical and acoustic parameters in his master thesis. By applying short-time Fourier transform to the recorded string signal and using the method introduced in \ref{sec:method-damping-param}, we calculated the damping parameters as $d_1 = 2.9390$ and $d_3 = -9.1\times10^{-4}$ with goodness-of-fit $R^2 = 0.74$ and $p = 2.84\times 10^{-7}$. 

Then we calculated the coefficients of the low-pass filter $H_{lp}$ based on the estimate of damping parameters. The coefficients of the dispersion filter are calculated based on the physical parameters of the 21st string of the \emph{guzheng}. The pitch of 21st string is D2 (73.42 Hz), therefore the total delay needed is $48000/73.42 = 653.77$ samples. Both the integer delay and the filters contribute to the total delay, and the fraction delay needed is $1.8566$. The coefficients of the fraction delay filter are calculated using Equ. (\ref{equ:Thiran-coef}) by substituting $D = D_{frac} = 1.8566$. All filter coefficients are listed in the Table \ref{tab:filter-coef}. 

The excitation function $\mathcal{F}_{e2}(t)$ is chosen to produce a discrete rectangular pulse of length 144 samples as the plucking time is chosen to be 3 milliseconds. The pluck point is set to be at the $1/7$ of the string on the right. Other physical parameters of the 21st string have been measured and listed in Table \ref{tab:physical-properties-Guzheng}. Also, a recorded string signal of the 21st string and the impulse response of the body are kindly provided by Deng et al. 

Our proposed model is implemented in MATLAB with the number of modes $\mu = 80$ and sampling rate $f_s = 48kHz$. The synthesised string signal is shown in Fig. \ref{fig:synthesised-string}. We then produced the sound of the \emph{guzheng} as the convolution of the synthesised string signal and the impulse response of the \emph{guzheng} body. Commuted synthesis can also be applied by convolving the impulse response with the excitation force $\mathcal{F}_e$.

\section{Discussion}
\label{sec:discussion}

\subsection{Accuracy}

Generally speaking, the accuracy of a sound synthesis system is evaluated by the similarity between the synthesised sound and the sound from the real instrument. The evaluation is subjective and aesthetic because it depends on the preference of timbre, musical experiences and so on. Subjective evaluation of the sound quality is worth studying and related experiments are left for future work. In this study, we mainly focus on the comparison on spectrum as the technical aspect of the model accuracy. But again, we by no means consider the spectrum comparison is the only way to prove accuracy and the quality of a synthesis method.


We compared our proposed method to other two state-of-art approaches. One is the ordinary Functional Transformation Method (FTM) as a modal synthesis method from directly from physical model. Another one is a Digital Waveguide model with the considerations of the damping and dispersion effect proposed by Trautmann and Rabenstein \cite{trautmann2002combining}. 

We synthesise the string signal of the \emph{guzheng} using Functional Transformation Method and Digital Waveguide. Comparing the spectrum of both synthesised string signals, our proposed method generally yields higher accuracy in the spectrum. The comparison of errors is shown in Table.\ref{tab:error-comparison}.

\begin{table}[!tb]
\scriptsize
	\centering
	\caption{The error comparison of each mode in the spectrum}
	\begin{tabular}{c|ccccc}
		\hline \hline
		Method           & 1st             & 2nd             & 3rd             & 4th             & 5th             \\
		\hline
		Proposed method  & -               & 0.0412          & \textbf{0.1061} & \textbf{0.0005} & 0.1831 \\
		FTM              & -               & \textbf{0.0360} & 0.9287         &  0.6568          & \textbf{0.0746} \\
		DWG model        & -               & 0.1931          & 0.4757          & 0.3140          & 0.2653          \\
		
		\hline \hline
		Method           & 6th             & 7th             & 8th             & 9th             & 10th            \\
		\hline
		Proposed method  & \textbf{0.1056} & 0.0529          & \textbf{0.0397} & 0.1408          & \textbf{0.0033} \\
		FTM              &   0.2156       &  \textbf{0.0392} &   0.1265        &    0.0684       &  0.1678\\
		DWG model        & 0.1279          & 0.1775          & 0.1148          & \textbf{0.0045} & 0.0889          \\
		
		\hline\hline
	\end{tabular}
	
	\label{tab:error-comparison}
\end{table}

\subsection{Performance}

We also measured the processing time of our proposed method with FTM and DWG model as two state-of-art approaches. The two comparison approaches are in the same parameter settings as in the accuracy evaluation. 

\begin{figure}[!tb]
	\centering
		\includegraphics[width=0.4\textwidth]{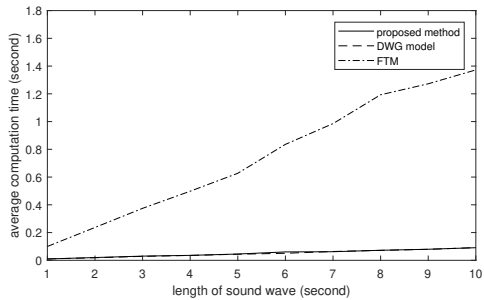}
		\caption{The performance comparison between our proposed method, DWG model and FTM.}
		\label{fig:performance}
	
\end{figure}

We synthesise the string signal for different lengths using all these three methods and record the processing time for each. When generating different length of synthesised signal, every method was evaluated 10 times to reduce the error. As shown in Fig. \ref{fig:performance}, the processing time is linear for every methods. The FTM model has a significant high computational complexity compared to our proposed method. Compared to DWG model, our proposed method takes almost the same time to produce the signal but requires a bit more computation, because of the  synthesis of the excitation signal. However, as the length of the excitation signal is usually fixed, the computation will not increase much and the slope of the two lines in Fig.\ref{fig:performance} are almost the same.




\section{Conclusion}
\label{sec:conclusion}

In this work, we proposed an efficient modal-based approach for the sound synthesis of the \emph{guzheng} which keeps the accuracy from the physical model and takes advantages from the use of DWG by simulating the propagation of the vibration on the string. We adopt a new design of the dispersion filter to be more applicable and practical. The proposed model is more flexible as it enables parametric control of the pluck. Our proposed model was implemented in MATLAB. By comparison with DWG model and FTM, our proposed model is proved to be acceptably accurate and efficient.

By now, only the 21st string of the \emph{guzheng} is synthesised because of limited time, while other strings can be easily simulated by similar models given particular physical parameters. The simulation of other strings and the modelling of non-linearities of the string \cite{trautmann2000sound} could also be done as future work.

\section{Acknowledgement}

We would like to express special thanks to Professor Zhengyue Yu and his formal master student Xiaowei Deng, who kindly provided lots of experiments data and continuously supported this work.

\vfill
\pagebreak


\bibliographystyle{IEEEtran}
\bibliography{ref}

\begin{thebibliography}{10}
\providecommand{\url}[1]{#1}
\csname url@samestyle\endcsname
\providecommand{\newblock}{\relax}
\providecommand{\bibinfo}[2]{#2}
\providecommand{\BIBentrySTDinterwordspacing}{\spaceskip=0pt\relax}
\providecommand{\BIBentryALTinterwordstretchfactor}{4}
\providecommand{\BIBentryALTinterwordspacing}{\spaceskip=\fontdimen2\font plus
\BIBentryALTinterwordstretchfactor\fontdimen3\font minus
  \fontdimen4\font\relax}
\providecommand{\BIBforeignlanguage}[2]{{%
\expandafter\ifx\csname l@#1\endcsname\relax
\typeout{** WARNING: IEEEtran.bst: No hyphenation pattern has been}%
\typeout{** loaded for the language `#1'. Using the pattern for}%
\typeout{** the default language instead.}%
\else
\language=\csname l@#1\endcsname
\fi
#2}}
\providecommand{\BIBdecl}{\relax}
\BIBdecl

\bibitem{Deng2015dynamic}
X.~Deng, Z.~Yu, W.~Yao, and M.~Chen, ``{D}ynamic analysis on {G}uzheng string
  vibration and bridge transmission,'' \emph{Journal of Vibration and Shock},
  vol.~34, no.~18, pp. 166--170, 2015.

\bibitem{Deng2015thesis}
X.~Deng, ``{A}nalysis of structure vibration and acoustic characteristics of
  chinese musical instrument {G}uzheng,'' Master's thesis, Shanghai Jiao Tong
  University, 2015.

\bibitem{Deng2016simulation}
X.~Deng, Z.~Yu, W.~Yao, and M.~Chen, ``{S}imulation analysis of vibro-acoustic
  characteristics of traditional {G}uzheng,'' \emph{Journal of Shanghai Jiao
  Tong University}, vol.~50, no.~2, pp. 300--305, 2016.

\bibitem{smith1993efficient}
J.~O. Smith, ``Efficient synthesis of stringed musical instruments,'' 1993.

\bibitem{smith2010physical}
------, \emph{Physical Audio Signal Processing: For Virtual Musical Instruments
  and Audio Effects}.\hskip 1em plus 0.5em minus 0.4em\relax
  http://ccrma.stanford.edu/\~jos/pasp/, Accessed 1st Mar 2018, online book,
  2010 edition.

\bibitem{valimaki1998development}
V.~V{\"a}lim{\"a}ki and T.~Tolonen, ``Development and calibration of a guitar
  synthesizer,'' \emph{Journal of the Audio Engineering Society}, vol.~46,
  no.~9, pp. 766--778, 1998.

\bibitem{lindroos2011parametric}
N.~Lindroos, H.~Penttinen, and V.~V{\"a}lim{\"a}ki, ``Parametric electric
  guitar synthesis,'' \emph{Computer Music Journal}, vol.~35, no.~3, pp.
  18--27, 2011.

\bibitem{gabrielli2013digital}
L.~Gabrielli, V.~V{\"a}lim{\"a}ki, H.~Penttinen, S.~Squartini, and S.~Bilbao,
  ``A digital waveguide-based approach for clavinet modeling and synthesis,''
  \emph{EURASIP Journal on Advances in Signal Processing}, vol. 2013, no.~1, p.
  103, 2013.

\bibitem{rabenstein2003digital}
R.~Rabenstein and L.~Trautmann, ``Digital sound synthesis of string instruments
  with the functional transformation method,'' \emph{Signal Processing},
  vol.~83, no.~8, pp. 1673--1688, 2003.

\bibitem{bank2010modal}
B.~Bank, S.~Zambon, and F.~Fontana, ``A modal-based real-time piano
  synthesizer,'' \emph{IEEE transactions on audio, speech, and language
  processing}, vol.~18, no.~4, pp. 809--821, 2010.

\bibitem{trautmann2002combining}
L.~Trautmann, B.~Bank, V.~Valimaki, and R.~Rabenstein, ``Combining digital
  waveguide and functional transformation methods for physical modeling of
  musical instruments,'' in \emph{Audio Engineering Society Conference: 22nd
  International Conference: Virtual, Synthetic, and Entertainment Audio}.\hskip
  1em plus 0.5em minus 0.4em\relax Audio Engineering Society, 2002.

\bibitem{erkut2002acoustical}
C.~Erkut, M.~Karjalainen, P.~Huang, and V.~V{\"a}lim{\"a}ki, ``Acoustical
  analysis and model-based sound synthesis of the kantele,'' \emph{The Journal
  of the Acoustical Society of America}, vol. 112, no.~4, pp. 1681--1691, 2002.

\bibitem{waltham2016acoustical}
C.~Waltham, Y.~Lan, and E.~Koster, ``An acoustical study of the qin,''
  \emph{The Journal of the Acoustical Society of America}, vol. 139, no.~4, pp.
  1592--1600, 2016.

\bibitem{penttinen2006model}
H.~Penttinen, J.~Pakarinen, V.~V{\"a}lim{\"a}ki, M.~Laurson, H.~Li, and
  M.~Leman, ``Model-based sound synthesis of the {G}uqin,'' \emph{The Journal
  of the Acoustical Society of America}, vol. 120, no.~6, pp. 4052--4063, 2006.

\bibitem{karjalainen1993analysis}
M.~Karjalainen, J.~Backman, and J.~Polkki, ``Analysis, modeling, and real-time
  sound synthesis of the kantele, a traditional finnish string instrument,'' in
  \emph{Acoustics, Speech, and Signal Processing, 1993. ICASSP-93., 1993 IEEE
  International Conference on}, vol.~1.\hskip 1em plus 0.5em minus 0.4em\relax
  IEEE, 1993, pp. 229--232.

\bibitem{trautmann2003digital}
L.~Trautmann and R.~Rabenstein, \emph{Digital sound synthesis by physical
  modeling using the functional transformation method}.\hskip 1em plus 0.5em
  minus 0.4em\relax Springer Science \& Business Media, 2003.

\bibitem{rauhala2006tunable}
J.~Rauhala and V.~Valimaki, ``Tunable dispersion filter design for piano
  synthesis,'' \emph{IEEE Signal Processing Letters}, vol.~13, no.~5, pp.
  253--256, 2006.

\bibitem{thiran1971recursive}
J.-P. Thiran, ``Recursive digital filters with maximally flat group delay,''
  \emph{IEEE Transactions on Circuit Theory}, vol.~18, no.~6, pp. 659--664,
  1971.

\bibitem{laakso1996splitting}
T.~Laakso, V.~Valimaki, M.~Karjalainen, and U.~Laine, ``Splitting the unit
  delay [{F}{I}{R}/all pass filters design],'' \emph{IEEE Signal Processing
  Magazine}, vol.~13, no.~1, pp. 30--60, 1996.

\bibitem{trautmann2000sound}
L.~Trautmann and R.~Rabenstein, ``Sound synthesis with tension modulated
  nonlinearities based on functional transformations,'' \emph{Acoustics and
  Music: Theory and Applications (AMTA), NE Mastorakis, Ed., Jamaica}, pp.
  444--449, 2000.

\end{thebibliography}
\end{document}